# Pertinent Information retrieval based on Possibilistic Bayesian network : origin and possibilistic perspective

Kamel Garrouch[1], Mohamed Nazih Omri[2], Bachir el Ayeb[2]
[1] ISSAT Sousse, TUNISIA – kamelg_2001@yahoo.fr
[2] FSM Monastir, TUNISIA – nazih.omri@ipeim.rnu.tn
[2] FSM Monastir, TUNISIA – ayeb_b@yahoo.com

*Abstract*—In this paper we present a synthesis of work performed on tow information retrieval models:

Bayesian network information retrieval model witch encode (in) dependence relation between terms and possibilistic network information retrieval model witch make use of necessity and possibility measures to represent the fuzziness of pertinence measure. It is known that the use of a general Bayesian network methodology as the basis for an IR system is difficult to tackle. The problem mainly appears because of the large number of variables involved and the computational efforts needed to both determine the relationships between variables and perform the inference processes. To resolve these problems, many models have been proposed such as BNR model. Generally, Bayesian network models doesn't consider the fuzziness of natural language in the relevance measure of a document to a given query and possibilistic models doesn't undertake the dependence relations between terms used to index documents. As a first solution we propose a hybridization of these two models in one that will undertake both the relationship between terms and the intrinsic fuzziness of natural language. We believe that the translation of Bayesian network model from the probabilistic framework to possibilistic one will allow a performance improvement of BNRM.

## I. INTRODUCTION

The field of information retrieval (IR) has been defined by Salton [10] as the subject concerned with the representation, storage, organization, and accessing of information items. The IR process consists in selecting among a large collection a set of documents that are relevant to a user's query.

Given a document collection, the first step to operate with an IRS, is to characterize the content of the document, task called indexing. It obtains a representation of each document in a suitable form to be managed by a computer. The result is a set of keywords or terms extracted from each text that should appropriately express the content of the document.

Because they are not equally important, these terms could be weighted to highlight their importance in the documents they belong, as well as in the whole collection. A weighted indexed document could be $D_j = \{(t_{1j}, w_{1j}), ..., (t_{kj}, w_{kj})\}$, where each $w_{ij}$ is the weight associated to the corresponding term. Usually, we use the weight known as tf/idf weight. In this case, the value associated to a term is computed multiplying the frequency of the term in that document (tf) by the inverse document frequency (idf) of the term in the collection.

When the indexing process has finished and the collection is ready to be used, a user interacts with the IRS by means of a query. That query is a description of the user's information need, and must be also indexed to produce a representation which can be handled by the system. The next step is the retrieval of those documents which are the most relevant to the query. The matching process is based on the search strategy implemented by the corresponding model. The result of this stage is a ranking of documents sorted by the proximity of each document to the query.

The set of retrieved documents in answer to a query does not usually correspond to the set of documents that are relevant to the user need. The relevance of a document to a query is usually interpreted by most of IR models, vector space, Boolean, probabilistic, as a score computed by summing the inner products of term weights in the documents and query representations. Whatever the used model, the response to a user need is a list of documents ranked according to a relevance value.

Generally most information retrieval model suffers from two drawbacks. The first one is related to the process of indexing and the second problem is related to relevance measure. These problems are caused by the fuzzy nature of natural language and by the semantic ambiguity of words.

Most model of IR doesn't consider these problems. To deal with the first problem Bayesian network model for IR has been proposed by de Campos [14] as solution. To deal with the second one, Brini [1] have proposed a IR model based on possibility theory.

Following these ideas, this paper is divided into the following sections: in Section 2, we introduce the Bayesian network background needed to understand the rest of the paper, we present its use in IR field and we explain the Bayesian Network Retrieval Model (BNRM). Section 3 presents the possibilistic model for information retrieval. Section 4 shows the conclusions, as well as future work that we plan to make in order to improve retrieval performance.

[1] This work is supported by the research unit PRINCE.

## II. BAYESIAN NETWORK

A Bayesian network is a directed acyclic graph (DAG), in which the nodes represent random variables and the arcs show causality, relevance or dependency relationships between them. The variables and their relationships comprise the qualitative knowledge stored in a Bayesian network. A second type of knowledge also stored in the DAG is known as quantitative. Associated with each node there is a set of conditional probability distributions, one for each possible combination of values that its parents can take. It establishes the strength of the relationships and is measured by means of probability distributions.

Bayesian network can be considered as an efficient representation of a joint probability distribution that takes into account the set of independence relationships represented in the graphical component of the model. In general terms, given a set of variables $\{X_1, \ldots, X_n\}$ and a Bayesian network G, the joint probability distribution in terms of local conditional probabilities is obtained as follows

$$P(X_1, \ldots, X_n) = \prod_{i=1}^{n} P(X_i | \pi(X_i))$$

Where $\pi(X_i)$ is any combination of the values of parent set of $X_i$, $\Pi(X_i)$, in the graph. If $X_i$ has no parents, then the set $\Pi(X_i)$ is empty, and therefore $P(X_i | \pi(X_i))$ is just $P(X_i)$.

Once completed, a Bayesian network can be used to derive the posterior probability distribution of one or more variables since we have observed the particular values for other variables in the network, or to update previous conclusions when new evidence reach the system.

The first important Bayesian network-based IR model was the inference network model, designed by Turtle and Croft [11]. It is composed, of two networks: the collection network and the query network. The former is composed of two types of nodes which represent documents and terms (concepts) symbolizing the index terms contained in the documents. The latter is built to represent the queries submitted to the system by means of query nodes and query concept nodes. The document network is fixed for a given document collection, and the query network is created each time that a user formulates a query. Once the probabilities have been assessed for each node, inference is carried out instantiating each document node, in turn. Therefore, the probability that the query is met given that a document has been observed in the collection is obtained. After all the propagation processes, the posterior probabilities are sorted in decreasing order, so the higher the probability the more relevant the document is [14].

The use of a general Bayesian network methodology as the basis for an IR system is difficult to tackle. The problem mainly appears because of the large number of variables involved and the computational efforts needed to both determine the relationships between variables and perform the inference processes. Nevertheless, an increasing effort has been made in the research of uncertain inference models for IR. These models consider the following two main simplifying restrictions in order to solve the above efficiency problem:

1- Fixed dependence relationships: the structure of the model, encoding the dependence relationships between variables, is fixed a priori, without considering any potential knowledge that might be mined from the collection.

2- Simplified estimation of probabilities: in order to avoid the large space necessary to store all the probabilities relevant to the process, it is assumed that those complex compound events will have been assigned zero probability values. With this assignment, these events can be discarded when inference tasks are performed.

Using the restrictions above, the probability of relevance of a given document only depends on the set of terms used to formulate the query and it can be computed without truly performing inference tasks, i.e. without propagating the evidences through the networks [14].

Based on the first simplification, many models have been proposed. The main difference between them is in the number of subnetworks, in the orientation of arcs and in the modelling of the (in)dependence relation between term's node. We are briefly going to review some works developed based on Bayesian networks.

IN the model proposed by [13] two different sets of nodes can be found : a set containing binary random variables representing the terms in the glossary from a given collection *(term subnetwork)*, and a second, corresponding also to binary random variables, but in this case related to the documents which belong to the collection *(document subnetwork)*. The orientation of arcs is as follow : the document nodes will only receive the arcs from term nodes and not from other document nodes. The relation between documents only occurs through the terms included in these documents.

Based on the fact that structured queries can be more expressive than their flat (natural language) query counterparts and that retrieval models that can evaluate structured queries have more potential to satisfy the user's information need, [9] address the problem of probability estimation in the inference network model and the problem of expressing structure in queries in a language modeling system by combining the two frameworks. The combination uses inference nets to express complex queries and language models to estimate the probabilities needed to evaluate those queries.

Many other works have been done in this field such as [2], [3], [4], [5], [6], [13], [15], [16], [17]. [12] explain briefly some of these works.

In short, most models use a fixed document subnetwork structure for a given collection. They do not take into account the particular dependence relationships between variables (terms and/or documents) that can be mined from the document collection. This is not the case for the BNR model.

**The BNR model**

In order to reduce these problems, de Campos [14] proposed a model called the Bayesian Network Retrieval Model. The objective was to obtain a model able to incorporate the most important dependence relationships in the collection, by the use of learning procedure.

The model, consider two sets of variables: Terms (T= {$T_i$, i=1,…,M} with M being the number of terms used to index the collection and documents D = {$D_j$, j=1,….,N), N being the total number of documents. The model is composed of two different layers of nodes: the term and the document layer. The former is used to incorporate the most important dependence relationships between the terms into the collection. Term to term dependences is represented by means of a polytree.

The relationship between a document and each of the terms by which it has been indexed is presented by the links between the tow layers. The relationships between documents are only present through the terms that index them. Thus documents are conditionally independent given the terms by which they have been indexed.

Fig. 1 shows an example of the final topology of the network

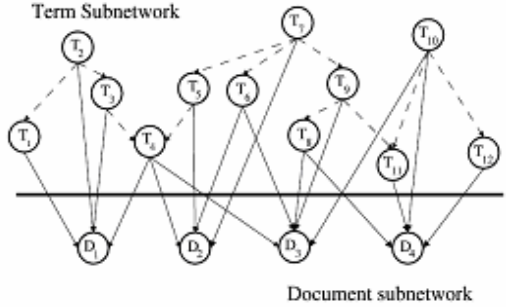

Fig.1. the Bayesian Network Retrieval Model

**Estimating the quantitative information**

Once the structure of the network has been created, the second step is to estimate the strength of the relationships represented. This process implies estimating a set of conditional probability distributions.

**Root term nodes**

Given a root node representing the variable $T_i$, the marginal probability of relevance, p(t), and the probability of being non-relevant, $p(\bar{t}_i)$ are defined by the mean of

p($t_i$) = 1/M and $p(\bar{t}_i)$ = 1- p($t_i$), with M being the number of terms in the collection.

**Non-root term nodes**

For each term node $T_i$, with parents $\Pi(T_i)$), we need to estimate a set of conditional probability distributions $p(T_i|\pi(T_i))$, one for each possible combination of values that the parents of a node $T_i$ can have, $\pi(T_i)$. The estimation is based on Jaccard similarity measure.

$$p(\bar{t}_i|\pi(T_i)) = \frac{n(\langle t_i, \pi(T_i)\rangle)}{n(\langle \bar{t}_i \rangle) + n(\pi(T_i)) - n(\langle \bar{t}_i, \pi(T_i)\rangle)}$$

In this formula, $p(\bar{t}_i|\pi(T_i))$ is initially estimated and later $p(t_i|\pi(T_i))$ is obtained by duality $p(t_i|\pi(T_i))$= 1- $p(\bar{t}_i|\pi(T_i))$

**Document nodes**

In this case, the probability $p(D_i|(\pi(D_i))$ must be estimated, i.e. the probability of a document node given the set of its parents (the nodes representing the terms by which it has been indexed). To estimate the probability matrices completely a probability functions, is used.

$$p(d_j|\pi(D_j)) = \sum_{T_i \in R_{\pi(D_j)}} w_{ij}$$

With **R** $\pi(D_j)$ being the set of terms that are relevant in $\pi(D_i)$ and the weights **$w_{ij}$** have to verify that **0≤$w_{ij}$** and $\sum_{T_i \in D_j} w_{ij}$ ≤1. So the more relevant terms in $\pi(D_i)$, the greater the probability of relevance of $D_j$

**Inference in the BNR**

On the BNR model, the query formulated by the user plays the role of a new piece of evidence provided to the system. The last aim is to obtain the probability of relevance of each document in the collection given a query. The terms from the query are instantiated to relevant in the network. This information will be propagated toward the document nodes, finally obtaining $p(D_j|Q), \forall D_j$.

The documents are presented to the user decreasingly sorted according to their corresponding probabilities of relevance.

To carry out the inference in an acceptable time the inference algorithm used is composed of two-stage approximate propagation:

1) Exact propagation in the term layer, obtaining $p(t_i|Q, \forall T_i$. Pearl's exact propagation algorithm is used in order to obtain the posterior probability of each term node. These probabilities can be computed in a polynomial time in an exact way.

2) Evaluation of a probability function in the document nodes, computing $p(D_j|Q), \forall D_j$ using the posterior probabilities obtained in the previous stage.

Therefore, the computation of $p(D_j|Q)$ can be carried out as follows:

$$p(d_j|Q) = \sum_{i=1}^{m_j} w_{ij}.p(t_i|Q)$$

## III. INTRODUCTION TO THE POSSIBILITY THEORY

Possibility theory treats uncertainty in a qualitative or quantitative way. Uncertainty in possibility theory is presented by a pair of dual measure of possibility and necessity, usually graded on the unit interval called possibilistic scale [7].

Possibility and necessity measures can take their values on qualitative ranges or on a numerical scale. This leads to two different forms of conditioning for qualitative and quantitative possibility measures, namely $\Pi(A \wedge B) = \Pi(A|B) * \Pi(B)$ holds taking $*$ as the minimum and the product respectively. We can then distinguish between qualitative and quantitative possibility theory. Qualitative possibility theory can be defined in purely ordinal settings, while quantitative possibility theory requires the use of numerical scale. Quantitative possibility theory was proposed as an approach to the representation of linguistic imprecision and then as a theory of uncertainty of its own **[8]**

Formally, a possibility distribution $\pi$ is a mapping form U to [0,1]. $\pi(u)$ evaluate the possibility that u is the actual value of some variable to witch $\pi$ is attached. $\pi(u) = 0$ means that u is impossible but $\pi(u) = 1$ only indicates a lack of surprise about u. A proposition A is evaluated by its degree of possibility $\Pi(A) = \max_{u \in A} \pi(u)$ and its degree of necessity (or certainty) $N(A) = 1 - \Pi(\overline{A})$ where $\overline{A}$ is the complement of A [1].

The importance of the theory of possibility stems from the fact that much of the information on which human decisions are based is possibilistic rather than probabilistic in nature. In particular, the intrinsic fuzziness of natural languages-which is a logical consequence of the necessity to express information in a summarized form-is, in the main, possibilistic in origin. Thus, when our main concern is with the meaning of information rather than with its measure the proper framework for information analysis is 'possibilistic' rather than probabilistic in nature [18]

**Possibilistic model for IR**

The model proposed by [1] is based on possibilistic directed networks, where relations between documents query and term nodes are quantified by possibility and necessity measures.

From a qualitative point of view, nodes in the graphical component represent query, index terms and documents and the graph reflects the (in)dependence relations existing between nodes.

A document $D_j$, take its values in the domain $\{d_j, \overline{d}_j\}$. The activation of a document node, i.e. $D_j = d_j$ (resp $\{\overline{d}_j\}$) means that a document is relevant or not. A query Q takes its values in the domain $\{q, \overline{q}\}$. The domain of an index term node $T_i$, is $\{t_i, \overline{t}_i\}$. ($T_i = t_i$) means a term $t_i$ is present in the object (document or query) and thus is representative of the object to a certain degree. A non-representative term, denoted by $\overline{t}_i$, is a term absent from the object. The proposed network architecture appears on Figure (2).

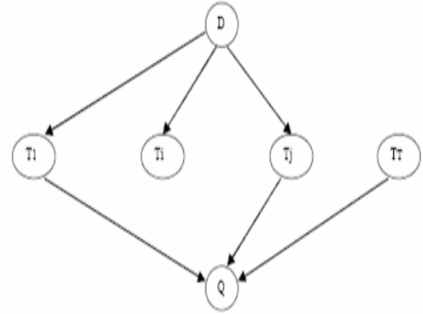

Fig2. The possibilistic information retrieval model

**Evaluation process**

In this model, the propagation process is similar to the probabilistic Bayesian propagation. The query evaluation consists in the propagation of new evidence through activated arcs to retrieve relevant documents. The model should be able to infer propositions like:

✓ It is plausible to a certain degree that the document is relevant for the user need, denoted by $\Pi(d_j | Q)$.

✓ It is almost certain (in possibilistic sense) that the document is relevant to the query, denoted by $N(d_j | Q)$.

The first kind of proposition is meant to eliminate irrelevant documents (weak plausibility). The second answer focuses attention on what looks very relevant.

Under a possibilistic approach, given the query, we are interested in retrieving necessarily or at least possibly relevant documents. Thus, the propagation process evaluates the following quantities: the degree of possibility and the degree of necessity:

$$\Pi(d_j | Q) = \frac{\Pi(Q \wedge d_j)}{\Pi(Q)}$$

$$N(d_j | Q) = 1 - \Pi(\overline{d}_j | Q)$$

Where

$$\Pi(\overline{d}_j | Q) = \frac{\Pi(Q \wedge \overline{d}_j)}{\Pi(Q)}$$

The possibility of Q is $\Pi(Q) = \max(\Pi(Q \wedge d_j), \Pi(Q \wedge \overline{d}_j))$

Given the model architecture, $(\Pi(Q \wedge d_j)$ is defined by

$$\max_{\theta}(\Pi(Q | \theta) \cdot \prod_{T_i \in T(Q) \wedge T(D_j)} \Pi(\theta_i | D_j) \cdot \Pi(D_j) \cdot \prod_{T_k \in T(Q) \setminus T(D_j)} \Pi(\theta_k)$$

for $\theta$ being the possible instances of the parent set of Q, $\theta_i$ is the instance of $T_i$ in $\theta$.

This is computed for $Dj \in \{d_j, \overline{d}_j\}$. Note that terms $T_i \in T(D_j) \setminus T(Q)$ are not involved in this computation.

The top retrieved documents are those having a necessary relevance value greater than 0, and the set of possibly relevant documents are retrieved as a second choice.

IV. CONCLUSION AND FUTURE WORK

In this paper we have presented tow information retrieval models: the first, BNR model encode the dependence relations between terms used to index documents. The second, PIR represent the relevance of a document to a query by necessity and possibility measures. PIR take into account the intrinsic fuzziness of the concept of relevance. However the dependence relation between terms is ignored. The hybridization of these models seems to be a good idea to deal with the fuzzy nature of the information retrieval process. We are currently studying this hybridization witch we didn't try out yet. We hope that an adaptation of both the algorithms used to learn the typologies and the quantitative knowledge of the network, and the inference technique to a possibilistic causal network adopted will show an improvement in the retrieval process.